# Fixing Inclusivity Bugs for Information Processing Styles and Learning Styles

Zoe Steine-Hanson, Claudia Hilderbrand, Lara Letaw, Jillian Emard, Christopher Perdriau, Christopher Mendez, Margaret Burnett, Anita Sarma
School of Electrical Engineering and Computer Science
Oregon State University
Corvallis, OR, USA
{steinehz, minic, letawl, emardj, perdriac, mendezc, burnett, anita.sarma}@oregonstate.edu

*Abstract*— Most software systems today do not support cognitive diversity. Further, because of differences in problem-solving styles that cluster by gender, software that poorly supports cognitive diversity can also embed gender biases. To help software professionals fix gender bias "bugs" related to people's problem-solving styles for information processing and learning of new software we collected inclusivity fixes from three sources. The first two are empirical studies we conducted: a heuristics-driven user study and a field research industry study. The third is data that we obtained about a before/after user study of inclusivity bugs. The resulting seven potential inclusivity fixes show how to debug software to be more inclusive for diverse problem-solving styles.

## I. INTRODUCTION

In recent years, diversity in computing has become prevalent; in fact, the current year's conference theme is "Diversity in Computing" [11]. One aspect of diversity that has not received much attention is diversity of *cognitive styles*. When software systems do not support cognitive diversity, people with some cognitive styles are forced to pay an additional "cognitive tax".

This paper considers cognitive inclusivity that relates to four cognitive styles from the GenderMag method [8]: (1) *comprehensive* information processing style, (2) *selective* information processing style, (3) learning by *process*, and (4) learning by *tinkering*. We refer to the first two styles using the category Information Processing Style (InfoProc style) and the second two styles using the category Learning Style.

Supporting cognitive diversity also helps support *gender* diversity, because the InfoProc and the Learning cognitive styles cluster differently for women than they do for men. Women more frequently favor comprehensive information processing and learning by process, whereas men more frequently favor selective information processing and learning by tinkering [1, 2, 3, 4, 5, 9, 10, 12, 13, 15, 17, 26, 33, 34, 39, 40].

Thus, when a software product does not support comprehensive information processing and process-oriented learning, it disproportionately disadvantages women. For example, research into gender diversity in Open Source Software (OSS) projects [31] reported gender biases in 73% of OSS newcomers' barriers, of which 71-72% were about Learning and 48% with InfoProc styles; other studies have also confirmed existence of gender biases in software [7, 18, 31, 42, 47]. We term these biases in software as "inclusivity bugs".

Fixing such inclusivity bugs requires knowing how cognitive styles differ and affect software use. For example, users with comprehensive InfoProc styles often seek relatively complete information upfront so they can plan their actions [28, 33, 34]. But, an excess of information can overwhelm users, especially users with selective InfoProc styles. Fixing these InfoProc inclusivity bugs means finding how to create interfaces that give comprehensive InfoProc users the amount of information they want, when they want it, and doing so without creating overwhelming interfaces.

Another way to address inclusivity bugs is by understanding the diversity in how individuals learn software new to them (e.g., what is the user's process to get to a solution, what does the user judge to be main vs. optional features). Some users approach this task by tinkering and exploring the software on their own, while others prefer to follow a more structured and guided path to understanding the software [3, 6, 15, 21, 43]. Supporting both of these learning styles means giving users room to tinker, but also giving users a process that doesn't require tinkering.

What fixes can potentially accomplish the goal of making inclusive software? To find out, we ran two empirical studies and obtained data from a third study, in which diverse design teams created design components aiming to address inclusivity bugs for InfoProc and Learning styles. We matched their work with evidence for solving InfoProc and/or Learning style inclusivity bugs, and triangulated these fixes with previous literature. Thus, our research question was:

*RQ: What are potential fixes to software inclusivity bugs for people's Information Processing styles and/or Learning styles?*

## II. BACKGROUND

### A. What is GenderMag?

Our investigation is in the context of the GenderMag method, [7, 9, 25, 31, 47], so we briefly summarize it here.

The GenderMag method involves using gender-associated personas, "Abi", the "Pats" and "Tim", to capture how people problem solve in different ways. The personas share five facets (styles), but each has their own facet values (Table I). To begin

5/7/19 9:05 PM

to fill the gap in creating inclusivity fixes, this paper focuses on two of them, Information Processing Style (InfoProc), and Learning Style.

The GenderMag personas and their facets are for helping software designers detect inclusivity bugs during a specialized Cognitive Walkthrough (CW) [49]. Several commercial and open source software teams have used GenderMag to detect inclusivity bugs in their software products [7, 18, 31, 42]. To use GenderMag, evaluators walk step-by-step through a software interface, asking the following modified set of CW questions from the perspective of one of the GenderMag personas:

SubgoalQ: Will <persona> have formed this subgoal as a step to their overall goal? (Yes/no/maybe, why, what facets are involved in your answer).
ActionQ1: Will <persona> know what to do at this step? (Yes/no/maybe, why, what facets...).
ActionQ2: If <persona> does the right thing, will s/he know s/he did the right thing and is making progress toward their goal? (Yes/no/maybe, why, what facets...).

An inclusivity bug is defined as an answer to one of the above questions that contains a "no" or "maybe" response that is tied to at least one of the facets.

*B. Information Processing Style (InfoProc Style)*

The InfoProc style facet explains how users like to gather information in software systems, either comprehensively or selectively [33, 34]. Individuals with comprehensive InfoProc styles prefer to gather a lot of information before acting (e.g., reading a whole page of documentation before making a change to some code) whereas those with selective InfoProc styles prefer to gather small bits of information and tend to act on these bits of information more frequently (e.g., reading the relevant parts of the documentation page and acting on it as they come across it). Both styles can get users to their goals, but because of gender biases in software, selective InfoProc styles tend to be supported more often in software [7].

A user's preference toward comprehensive InfoProc does not mean that user has superhuman patience. Although users with this cognitive style prefer to fully scope out a problem before beginning to solve it, they may abandon their efforts if doing so is too cumbersome, time-consuming, nonsensical, etc. Thus, adding more information does not alone accommodate comprehensive information processors, and may hinder them. As previous researchers have suggested, what matters is *how much* and *when* information is provided and in what form [21, 28]. We discuss remedies for these concerns in Section V.

*C. Learning Style*

The Learning Style facet, which focuses on tinkering vs. process oriented learning styles, describes ways users approach learning how software works [3]. Some users prefer to learn about new software in process-oriented ways (e.g., tutorials that show the steps of bringing different features together), while others prefer to tinker and explore, constructing their own understanding of the software (e.g., trying out different options and backtracking if needed). Tinkering Learning styles tend to be better supported in software [7].

Analogous to comprehensive InfoProc, users who learn by process do not find all processes equally usable. If a process is overly long, complex, convoluted, etc., it may still present a usability barrier. Some research has shown possible software fixes to support process oriented learners [22, 28]. Here we discuss specific techniques to support diverse Learning styles from our datasets in Section VI.

III. INTRODUCING THE GENDERMAG HEURISTIC EVALUATION

In this investigation, we introduce a new variant of the GenderMag method: the GenderMag Heuristics. The GenderMag Heuristics are designed to enable software developers and designers, of varying experience levels, to find and create fixes to inclusivity bugs, much like Nielsen's Heuristics [35], but geared towards gender inclusivity. The heuristics are structured so that each problem-solving facet contains a brief description of the facet, details on how that facet influences each persona's actions, and an actionable "take away", which tells the heuristic evaluator how to evaluate their user interface for the whole spectrum of users for that facet (Table II).

IV. EMPIRICAL METHODS

To investigate evidence-based potential fixes for InfoProc and Learning style bugs, we ran two studies and obtained data from a third. We triangulated fixes across the studies, and with fixes from previous literature. The studies are briefly described in Table III.

*A. Study 1: Heuristics-Driven User Study*

To collect a variety of potential InfoProc and/or Learning Style inclusive fixes, Study 1 included 3 steps. For Steps 1-2, we recruited 18 novice HCI students (novice UXers) from an HCI class at a West Coast University (anonymized according to IRB guidelines). To complete Step 3, we recruited 6 user experience professionals (UXperts) from industry. Participants participated voluntarily, without pay. The steps are detailed below.

Step 1 (novice UXers): We gave the novice UXers a set of five inclusivity bugs that a prototype's original designers had identified (prior to the study) using the GenderMag method. The novice UXers had one week to individually develop potential fixes for each inclusivity bug using the GenderMag Heuristics (Section III; full assignment Fig. 5 in supplemental document).

Step 2 (UXer teams): We randomly assigned the novice UXers to teams of three. Each team met during a one-hour class period to combine, further develop, and/or further their individual

TABLE I. A SUMMARY OF THE FACET VALUES FOR EACH PERSONA. THIS PAPER FOCUSES ON INFOPROC STYLES AND LEARNING STYLES FACETS (IN GREY).

| | Abi | Tim |
|---|---|---|
| **Motivations** for using technology | Wants what the technology can accomplish | Technology is a source of fun |
| **Computer Self-Efficacy** about using unfamiliar technology | Low compared to peer group | High compared to peer group |
| **Attitude towards Risk** when using technology | Risk-Averse | Risk-Tolerant |
| **InfoProc Style** for gathering information to solve problems | Comprehensive | Selective |
| **Learning Styles** for learning new technology | Process-oriented | Tinkerer (sometimes to excess) |

results [41], producing a final evaluation and set of potential fixes.

Step 3 (UXpert reviews): We sent these potential fixes to the UXperts for their professional opinions. We asked the UXperts, who had previous experience with GenderMag (Table IV) to rate the potential fixes and, if needed, suggest refinements. Two UXperts reviewed each UXer team's potential fixes using a common rating rubric, which contained a 1-5 Likert scale on usability for each GenderMag persona (Fig. 3 in supplemental documents).

Step 4: We used the results of the UXperts' evaluations as a metric for filtering the novice UXers' potential fixes. We considered it a UXpert disagreement if they gave opposite responses to the Likert scale evaluation (i.e., UX1 answered "Unlikely" and UX2 answered "Extremely Likely"), and discarded these potential fixes. The remaining potential fixes that worked for at least 2 of the 3 personas, or had a concrete suggestion from the UXperts, were kept in the dataset. We then analyzed these novice UXer potential fixes and selected 7 that related to an InfoProc and/or Learning style inclusivity bug or the heuristic.

### B. Study 2: Industry Study

Study 2 was an Action Research study. Action Research is a type of long-term field research in which the researchers and participants jointly work together to not only investigate but also improve the topic under study [24, 30, 46]. Four professional software teams—two at a west coast university and two from two different companies—used the GenderMag method to find and address inclusivity bugs in their software.

The members of the four teams in this study included software and UX professionals, site admins, and marketing experts from a west coast University and two companies. Researchers introduced the teams to the GenderMag method during a pre-meeting and helped them run their first GenderMag session. Some teams held more researcher-assisted sessions using the script from the GenderMag kit, and some teams ran their own GenderMag sessions using the full GenderMag kit [8].

Researchers collected data of multiple types from participating teams: GenderMag forms (Fig. 2 in supplemental documents) filled out by teams during their GenderMag session(s), audio recordings (with transcriptions) of GenderMag sessions, the personas customized by teams (Fig. 1 in supplemental documents), notes from observing researchers (i.e., from GenderMag sessions, follow-up meetings, and post-interviews), team artifacts (e.g., screenshots, design mockups), and team communications (e.g., emails, social media activity). Semi-structured post-interviews were conducted when possible to gain additional data about the teams' potential fixes.

TABLE II. THE GENDERMAG HEURISTICS (INFOPROC AND LEARNING STYLES IN GREY)

| | Heuristic | Explanation |
|---|---|---|
| InfoProc | Let people gather as much information as they want, and no more than they want. [17, 32, 33, 39, 50] | People like to gather different amounts of information to solve problems:<br>Abi and Pat gather and read everything comprehensively before acting on the information.<br>Tim likes to delve into the first option and pursue it, backtracking if need be. |
| Learning Style | Provide a path through the task for process-oriented learners, and for tinkerers, encourage mindful tinkering (e.g., slow them down with an extra click), so that it is not so addictive. [6, 12, 15, 26, 40] | People learn software in different ways:<br>Abi learns better through process-oriented learning; (e.g., recipes, not just individual features).<br>Tim learns by tinkering (i.e., trying out new features) but sometimes he tinkers addictively and gets distracted by it.<br>Pat learns by trying out new features, but does so mindfully, reflecting on each step. |
| Motivations | Make clear what a new feature does, and why someone would use it, while also keeping familiar features available. [5, 6, 13, 20, 26, 40, 44] | People have different motivations for using technology:<br>Abi uses technology only as needed for her task. Abi prefers familiar and comfortable features to keep focused on her primary task.<br>Tim likes using technology to learn what new features can help him accomplish.<br>Pat is like Abi in some situations and like Tim in others. |
| Computer Self Efficacy | Make available ALL of (1) familiar features, (2) undo/redo, and (3) ways to try out different approaches, to support ALL self-efficacy levels. [5, 6, 14, 23, 27, 36, 37, 38, 45] | People have different amounts of self-efficacy (self-confidence) in using unfamiliar technology:<br>Abi has low self-efficacy about unfamiliar computing tasks. If problems arise with the technology, Abi often blames herself. This affects whether and how Abi will persevere.<br>Tim has high self-efficacy with technology. If problems arise with his technology, he usually blames the technology. He sometimes tries numerous ways of trying to address the problem before giving up.<br>Pat has medium self-efficacy with technology. If problems arise with his/her technology, s/he keeps trying for quite awhile. |
| Attitude Toward Risk | Make available why someone should use the feature (benefits) and how much effort it will take (cost); doing so supports decision making no matter their attitude toward risk. [16, 19, 48] | People tolerate different levels of risk (e.g., possibility of wasting a lot of time) when using technology:<br>Abi and Pat, who rarely have spare time, like familiar features because these don't require learning, and are predictable about the benefits and costs of using them.<br>Tim is risk tolerant and is ok with exploring new features, and sometimes enjoys it. |

TABLE III. THE STUDIES AND DATASET TEAMS AND THE APPLICATION THEY EVALUATED. TEAMS WITH * BY THEIR NAME ARE FROM A WEST COAST UNIVERSITY.

| Study/ Dataset | Study Description | Team | Application |
|---|---|---|---|
| Study 1: Heuristics-Driven User Study | Novice UXers ran heuristic evaluation and UX experts evaluated their redesigns. | Team 1A, 1B, 1C, 1D, 1E, & 1F | Mobile app used for tracking employment hours |
| Study 2: Industry Study | University and industry teams ran GenderMag sessions on various applications and recorded their results. | *Team 2L | University library applications |
| | | *Team 2W | Web app that was a data content template for end users using Drupal 8 |
| | | Team 2P | Web based interface for visual sorting with a deep learning back end |
| | | Team 2N | An IT-support product for end users |
| Dataset X: Before/ After User Study | Company X ran a user study before and after redesigning to address inclusivity bugs found using | Team XO | Academic search engine |

We included the following data in our analysis: filled-out GenderMag forms, session transcriptions, and post-interview responses. Post-interview questions (Fig. 4 in supplemental documents) were analyzed in cases when GenderMag forms and session transcripts were not available (2 out of 4 teams). When analyzing a GenderMag session transcript, we marked where team members gave suggestions for fixes to bugs in the interface, and then used the team's GenderMag forms to tie these suggestions back to the facets that caused the bug. For post-interview transcripts, we marked lines where a team mentioned potential fixes they made to their interface as a result of a GenderMag session.

We used the above data about these professional teams' potential fixes to triangulate with the potential inclusivity fixes generated by Study 1.

*C. Dataset X: Before/After User Study*

We obtained our third data source, Dataset X, from the authors of a Before vs. After laboratory study comparing the empirical results of a product's "before GenderMag" version against the same product after the product's owner made GenderMag-inspired potential fixes [47]. With permission from the study's authors, we analyzed their potential fixes to 5 (out of the 6) bugs they investigated—namely, the five that included an InfoProc or Learning style bug. These data contributed empirical evidence of the validity of some of the potential fixes from Study 1 and 2.

## V. INCLUSIVITY FIXES TO SUPPORT DIVERSE INFORMATION PROCESSING STYLES

*A. When to Present?*

Supporting diverse InfoProc styles means giving users the information they want, *when they want it*. One way teams attempted this was by combining context-sensitive help buttons (right where the action happens) with context-free help buttons (away from the immediate action to facilitate recall). Teams used these help widgets to reach users with comprehensive InfoProc style, while also doing no harm to selective InfoProc users. However, to support both styles, teams had to do more than simply "add help buttons", as we will see.

Team 1D created a potential InfoProc-inclusive fix to support both Abi's and Tim's information needs with context-sensitive help. The inclusivity bug was a design inconsistency in the

TABLE IV. UX EXPERT DEMOGRAPHIC DATA

| UX ID # | Gender | Age | UX Professional Experience | GenderMag Familiarity |
|---|---|---|---|---|
| UX1 | Woman | 21-30 | < 1 Year | Extremely familiar |
| UX2 | Woman | 21-30 | >= 5 Years | Moderately familiar |
| UX3 | Man | 31-40 | >= 5 Years | Extremely familiar |
| UX4 | Woman | 41-50 | >= 10 Years | Moderately familiar |
| UX5 | Woman | 21-30 | >= 5 Years | Slightly familiar |
| UX6 | Woman | 31-40 | >= 10 Years | Extremely familiar |

mobile app prototype: when users clicked a button labeled "Timesheet" (Fig. 1(A)), they were led to a page titled "Schedule". GenderMag evaluators thought this was an inclusivity bug because:

*GM Evaluation Bug #1: "the label on the top of the page says 'Schedule' whereas the user's goal was to open the timesheet. <Abi> may not realize that the schedule and timesheet are the same thing."*

To address this bug, Team 1D made two changes. First, they updated the title of the second page to match the button on the first page (titling it "Timesheet"). Second, they added context-sensitive help next to the Timesheet button (and other features of interest) in the user interface, which were designed to be in the form of clickable question marks. When clicked, the buttons would allow users to "gain additional information + help" about the (adjacent) feature. UXperts thought this potential fix was good, but that the screen needed fewer help buttons, to accommodate both Abi's and Tim's InfoProc style:

*UX1 Team 1D Bug #1: "Abi, due to her comprehensive information gathering style, would like the addition of the '?' icons… <but> Tim... will only look at the '?' buttons when he feels he really needs them… there is a chance he may accidentally click them instead of the buttons."*

The revised number and placement of context-sensitive buttons allowed Abi to learn more about where the Timesheet button leads, while also allowing Tim to ignore it (Fig. 1).

When users need help with something that's not on-screen, context-*free* help buttons can serve as InfoProc-inclusive design components as seen in [29]. During their development of the Idea Garden, a tool to help people learn how to program, the researchers found that context-free help buttons (e.g., help accessible from a toolbar), in addition to context-*sensitive* help buttons, were supportive of both comprehensive and selective InfoProc styles, because users could gather more information at more times when they needed it (further strengthening evidence of the benefits of context-sensitive help buttons).

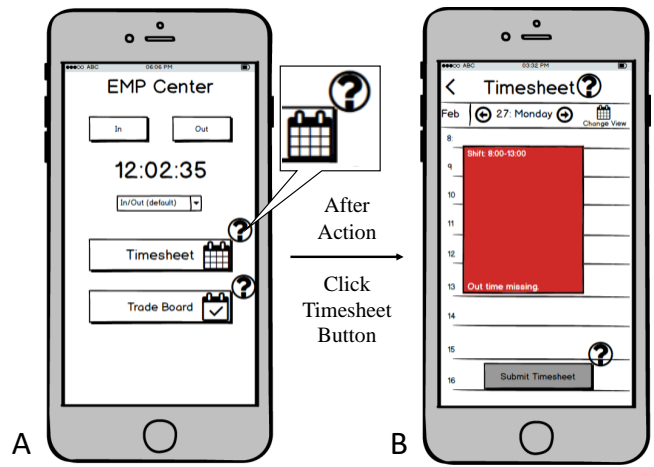

Fig. 1. The original interface looked the same, but included no context sensitive '?' buttons (circle button with ?, see callout). (A) The starting page for the application. (B) Page after clicking "Timesheet" button. The UXperts suggested that Team 1D's potenial fix needed to include '?' buttons only in problematic places instead of everywhere in order be useful to comprehensive InfoProc, without being cumbersome to others.

*B. How to Provide?*

Supporting both InfoProc styles means not only finding the right time to present information, but also knowing *how to provide* that information. To do so, some teams presented specific, but flexible information to the user. Team 2P used tooltips in their software to do this, noting that:

*Team 2P Interview: "tooltips are a really easy quick way to help **a little bit**" (emphasis added).*

These short tooltips can be useful for users with selective InfoProc, since the information is specific and to the point, but does not provide the complete information that comprehensive InfoProc users prefer. For teams to support both comprehensive and selective users, they had to make their tooltips expandable and pinnable.

Expandable/pinnable tooltips go beyond "a little bit of help", giving users the option to gather more information by expanding the tooltip, and the option to "pin" the tip in place to keep it on-screen. In a previous study, researchers supported both InfoProc styles via expandable/pinnable tooltips [28] (Fig. 2). Selective InfoProc users could focus on the short and specific information, whereas comprehensive InfoProc users could expand the tooltip for more information [28].

Team 1D aimed to support InfoProc styles in another way, by including specific but flexible information about *relevant data*. In one bug in the employee application in Study 1, Abi needed to check her remaining shifts before submitting the timesheet (Fig. 3 (A)), but at this point in the interface she could only see the shift she had just edited:

*GM Evaluation Bug #4: "If they are familiar with <timesheet applications> they should easily know how to go through and check the remaining days. Otherwise, they may experience some confusion".*

To address this inclusivity bug, Team 1D added a 'change view' button so users could see more or fewer shifts at once:

*Team 1D Bug #4: "<change view> button lets <users> access month view, or ... week view"* (Fig. 3 (B & C)).

The potential fix aimed to help the mobile app follow the InfoProc heuristic by providing comprehensive InfoProc users a way to get more information so they could feel comfortable with the task, while still letting selective InfoProc users only see specific bits of information. Comprehensive InfoProc users, especially users unfamiliar with the type of application, could use "change view" to scan through the whole month of shifts, getting a sense of the application and what needs to be done, before going back and modifying their shifts.

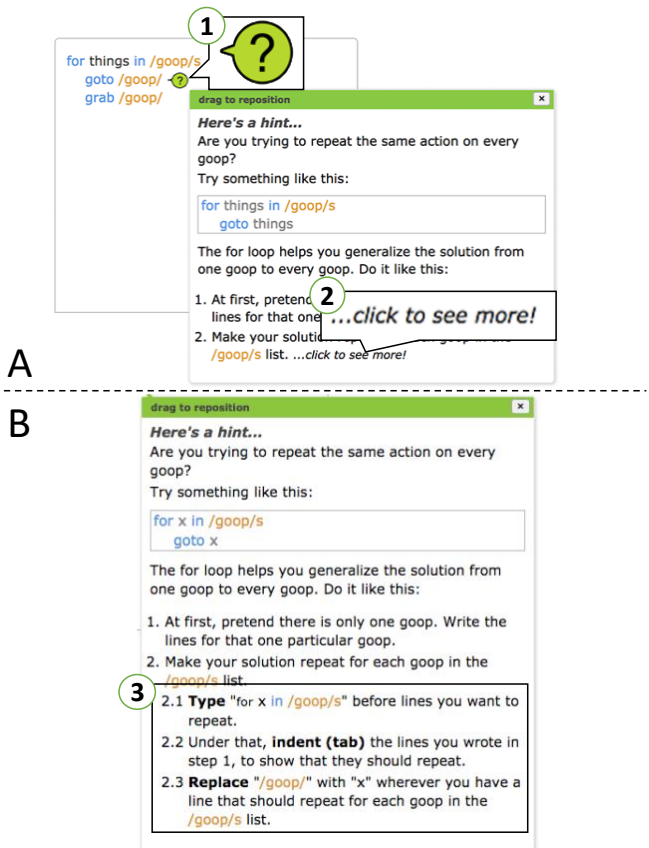

Fig. 2. An example of the types of expandable/pinnable tooltips used in [28]. (A) First view of the tooltip, which showed short and specific information for selective users. 1) Users opened the tooltips with the green '?' button. 2) Users could "click to see more" to expand the tooltip. (B) Expanded tooltip. The information at 3) shows comprehensive users more information about the programming topic.

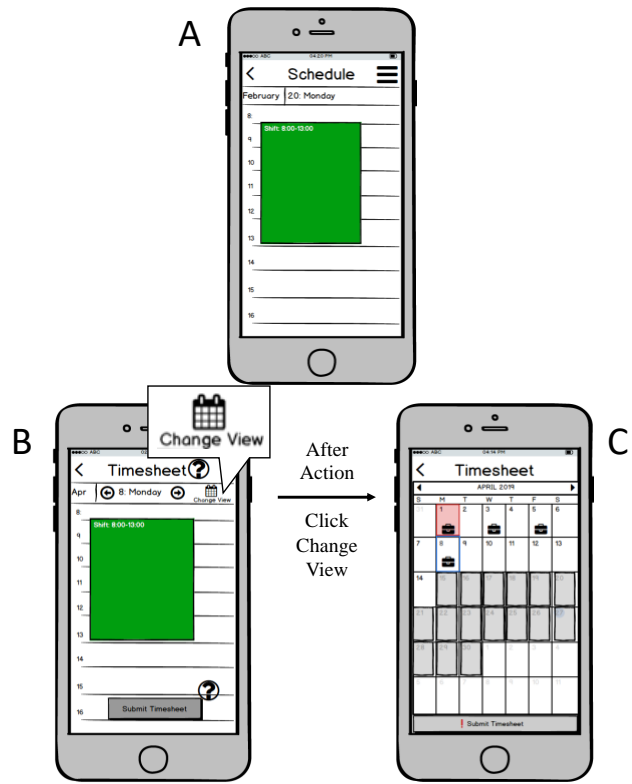

Fig. 3. (A) Screenshot of the prototype when developers decided the next action would be to go through the remaining days, but Abi might get confused due to her cognitive styles not being supported by the software. (B & C) Team 1D's final potential fix after UXpert feedback, includes a change view button (see callout) that lets the user decide whether to see more of the relevant information. (B) Day view of the shifts for selective InfoProc. (C) Month view of the shift for comprehensive InfoProc. Briefcases indicate a shift on that day and red days mean there is an issue. Days in grey are not in the current pay period.

The UXperts thought the potential fix would improve the inclusiveness of the mobile app, saying,

*UX1 Team 1D Bug #4:* "*the 'change view' option is smart*".

However, the UXperts also wanted more details about the calendar view itself, suggesting,

*UX1 Team 1D Bug #4:* "*some icon within each day <in the calendar view> that has a shift ... and then highlight whatever day the user has most recently selected*" (Fig. 3 (B & C)).

*Summary*: These potential fixes show that for teams to make software inclusive for InfoProc styles, teams needed to know the right time to give people help and how to design in flexibility of the *information*. Table V summarizes the potential fixes that teams made to support InfoProc styles with triangulation from previous research.

## VI. INCLUSIVITY FIXES TO SUPPORT DIVERSE LEARNING STYLES

### A. What's the Process?

To support process-oriented learners, some teams used a step-by-step formula (Fig. 4) to provide an *overview of the process*. Team XO found that the process of claiming an authorship in the academic search interface was unclear as it comprised steps split across multiple pages without indicating this. For one step in the process, a user needed to look over the papers on the screen to claim the right papers and Team XO thought this would cause an issue for Abi:

*Team XO Bug #5:* "*Abi might feel confused at this step... and not prefer tinkering*".

They recognized that, because of her Learning Style, Abi would want more information, but wouldn't tinker to get it. They addressed the bug by including a Step-by-Step formula that showed the steps of the authorship-claiming process, and which step Abi was on (Fig. 4).

### B. Am I Making Progress?

Only showing the overview of the process was often not enough, teams also showed users that *they were making progress towards their goals*. To do this, Team XO changed the academic search interface to clarify that the user was making progress. Because the original interface had the "claim authorship" button on both screens, Team XO thought that Abi would not know that she is making progress:

*Team XO Bug #5:* "*there's no feedback and instructions on what Abi should do next (learning)*".

The team recognized the Learning Style bug and suggested

*Team XO Bug#5 Solution:* "*...help process-oriented learners... instead of keeping the button the same for two steps ... update the button according to the process to make Abi know which step she is in*".

Their potential fix showed the user they were making progress: they modified the static "claim authorship" button to a series of buttons: "claim authorship"; "cancel" and "next". The label "Are any of these you?" stayed the same (giving users the context of the process step), but the button labels changed (helping them know they were progressing), as shown in Fig. 5 (B).

### C. Can I Move Forward?

Another way teams gave users a sense of process was by *constraining the next step* until the user completed their current step. Teams 1A and 1D addressed diverse Learning styles with their potential fix to the bug in the timesheet application (Fig. 6(A)), adding a 'submit' button to the timesheet page and greying it out until all shifts (inputs) were error free. Their potential fix aimed to show process-oriented learners what to do before proceeding, and to save tinkerers from having to backtrack if they clicked the 'submit' button too early. UXperts

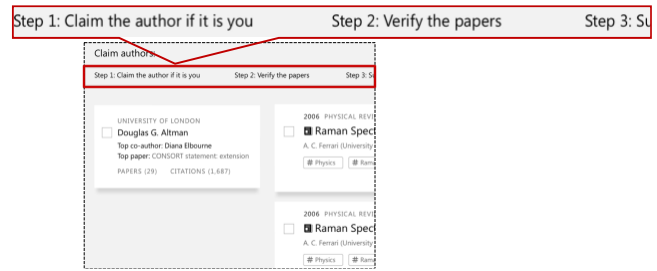

Fig. 4. Team XO added step-by-step progress formula at the top to help process oriented learners understand the whole process of claiming an authorship [47].

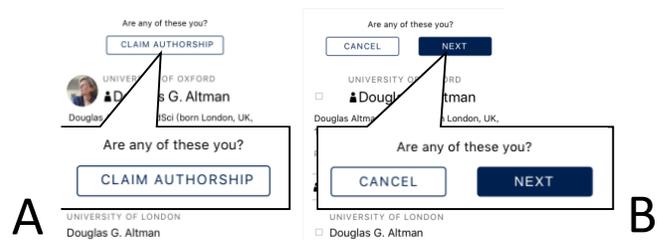

Fig. 5. (A) The Before version of the interface with the static "claim authorship" button that made it unclear to process oriented learners where they were in the process. (B) The After version of the interface with dynamic buttons that show process oriented learners they've made progress.

TABLE V. THE THREE POTENTIAL INFOPROC STYLE FIXES FROM TEAMS ACROSS THE THREE DATASETS. THE LAST COLUMN SHOWS EVIDENCE FOR THE POTENTIAL FIX (EITHER FROM DATASET X USER STUDY OR PREVIOUS LITERATURE). INFOPROC, FOLLOWED BY SYMBOLS, SHOWS THE RESULTS FROM DATASET X. THE FIRST SYMBOL IS FOR COMPREHENSIVE AND THE SECOND IS FOR SELECTIVE.

+ MEANS THAT THOSE USERS SAW AN IMPROVEMENT IN AFTER VERSION.
* MEANS THAT THOSE USERS HAD NO PROBLEMS IN EITHER VERSION.

| Description of potential fixes | Helps InfoProc? | | Instances | Evidence (Either literature o Empirical) |
| --- | --- | --- | --- | --- |
| | Compre-hensive | Selective | | |
| When to Present - Help context | Yes | neutral | Study 1: Team 1D Study 2: Team 2P | [29] |
| How to provide - Specific and Flexible: Expandable Tooltips | Yes | yes | Study 2: Team 2P | [28] |
| How to provide - Specific and Flexible: Multiple views of data | Yes | yes | Study 1: Team 1D Dataset X: Iss 1&2 | InfoProc + * [22] |

liked this potential fix, and suggested another fix to support process-oriented learners:

*UX4 Team 1A Bug #5: "Put<ing> some information near the disabled button about # of items with a problem"* See Fig. 6 (D2) for details.

The suggestion from the UXperts clarified the process even further by telling process-oriented learners why they could not proceed, and how many steps it would take to move on.

*D. What To-Do Next?*

Teams also clarified the process by showing *what to-do within a step in the process* with to-do lists. Recall the bug from Section V(B), where Abi needed to check her remaining shifts before submitting the timesheet (Fig. 3 (A)), but at this point in the interface she could only see the shift she had just edited. To address this bug, Team 1A designed a 'list view' of remaining days, which included a (!) notification on days with shift errors. These (!) notifications created an implicit to-do list for the user, telling them what shifts they still needed to fix. However, UXperts thought tinkerers might ignore the (!) button, so they suggested emphasizing shifts with errors using a red (!) notification and red text (Fig. 7).

The to-do list in this example was implicit, but explicit to-do lists can be just as helpful in highlighting the steps remaining in the process. For example, Team 2L used explicit checklists to support process-oriented users:

*Team 2L GM Session 2: "it's as if this [domain specific requirement] needs to be at the top and the requirements be a more prominent ... workflow-oriented thing, like a checklist".*

*Summary*: With their potential Learning Style inclusivity fixes, teams found that clarifying all steps, even the smallest ones, in the workflow can better support process-oriented learners, while also helping tinkerers avoid errors. A summary of these potential fixes from the teams and triangulation from previous literature is in Table VI.

VII. DISCUSSION: INCLUSIVITY IS IN THE NUANCES

On the surface, it may seem that many of the potential fixes the teams recommended are already known as good HCI practices. However, designing for inclusivity lies in the nuances. The nuances are not just in the fixes (e.g., expandable/pinnable tooltips), but also in how (e.g., on demand) and where (e.g., context-sensitive vs. context-free) they are applied that help cognitive diversity.

*Helping, but not overwhelming*: Helping comprehensive InfoProc users requires giving them information about specific components, but providing excessive detail about each and

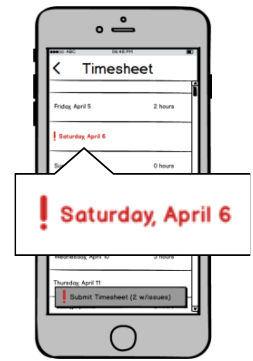

Fig. 7. The potential fix, with UXpert feedback, for Bug #4. Displays the remaining days as a list, with red '!' notifications and text for days with problems.

TABLE VI. FOUR POTENTIAL LEARNING STYLE FIXES. THE LAST COLUMN SHOWS EVIDENCE FOR THE POTENTIAL FIX (EITHER FROM DATASET X, USER STUDY, OR PREVIOUS LITERATURE). LEARNING, FOLLOWED BY SYMBOLS, SHOWS THE RESULTS FROM DATASET X. THE FIRST SYMBOL IS FOR PROCESS ORIENTED AND THE SECOND IS FOR TINKERERS.

+ MEANS THAT THOSE USERS SAW AN IMPROVEMENT IN AFTER VERSION.
* MEANS THAT THOSE USERS HAD NO PROBLEM IN EITHER VERSION.

| Description of the Potential Fix | Helps Learning? | | Instances | Evidence |
|---|---|---|---|---|
| | *Process* | *Tinker* | | |
| What's the Process? | yes | neutral | Dataset X: Iss6 Study 2: Team 2L | Learning * * |
| Am I Making Progress? | yes | neutral | Study 1: Teams 1A, 1C & 1E Dataset X: Iss5 Study 2: Teams 2N & 2W | Learning + * |
| Can I Move Forward? | yes | yes | Study 1: Teams 1A & 1D | |
| What To-Do Next? | yes | neutral | Study 1: Team 1A Study 2: Team 2L | [21] |

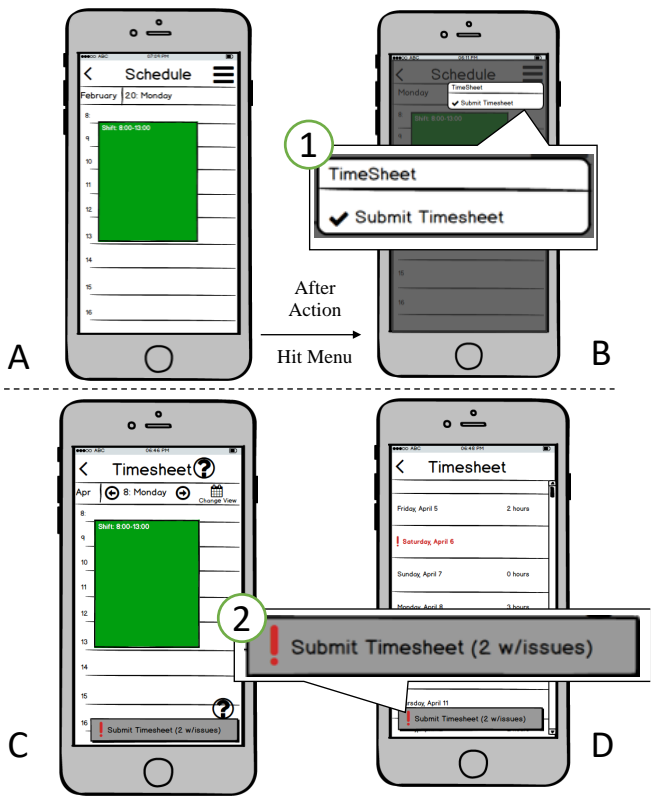

Fig. 6. (A&B) Screenshots of Bug #5. Abi needs to click the menu button ☰ (top right corner) to reach the submit timesheet button (1), but would not know to do this because of her Learning style facet. (C&D) Potential fixes with UXpert feedback from Team 1A and Team 1D, respectively, which included a greyed out submit button (2), and added information near the 'submit timesheet' button on how many shifts still have an issue.

every component of the interface can backfire. Abi likes to have comprehensive information before starting a task, but detailed information about every component in an interface can be overwhelming, confusing, and/or annoying. Too much help can also hinder Tim, as we saw from the Uxpert-suggested potential fix (Fig. 1), which used fewer help buttons to avoid being cumbersome.

Teams managed this tension between the amounts of information to provide by being mindful about: (1) *when* and *where* they made help available, using context-free and context-sensitive help buttons, and (2) *how* they made help available, giving users the flexibility to decide when and how much information they received. These nuances aimed to enable users to obtain as much information as they need, in the moment/context they need it, without being overwhelmed or underwhelmed.

*Not losing sight of Tim*: Some teams became so caught up in the lack of support they found for Abi, they sometimes lost sight of the goal of supporting *all* cognitive styles, including Tim's. Doing so enables support for not only more users, but also more situational needs, because a user's cognitive styles can vary from situation to situation. As we saw in Team 1D's use of context sensitive help buttons, UXperts thought comprehensive users might prefer buttons in all helpful places, but recognized that too many may harm other users' ability to use the software. The key was finding ways to support both styles simultaneously.

One way in which teams supported both learning styles was with to-do lists that showed users sequences of steps/actions they could take. The sequence supported process-oriented learners' understanding of the workflow, while also giving space for users to tinker and explore.

*More than the sum of its parts*? Since each cognitive style influences others and how people problem solve, potential fixes to bugs raised by the InfoProc and Learning style facets, may also support other cognitive styles. For example, Team 2P used tooltips to support Abi's InfoProc style, but expandable/pinnable tooltips may also support Abi's and Tim's Motivations. Expandable/pinnable tooltips can highlight the reason to use a feature and how it relates to the users' goal, thus supporting Abi. These tooltips can also provide a quick way to learn about different feature's functionality, thus supporting Tim's motivation to learn all the functionality of a software. The intertwined nature of the cognitive styles show that effectively supporting a particular facet can help make the software inclusive to other facets too.

Finally, the potential fixes from our data show that teams had to consider fixing software at different levels to make the whole better. For example, teams used context sensitive help to support users during the action, but to make the whole software inclusive, teams had to support users outside of the action as well (context free help). Similarly, teams needed to support users' understanding of the process at multiple levels. Teams clarified the immediate next steps with to-do lists, but also gave user context of the overall process with step-by-step formulas. Integrating potential inclusive fixes at all levels helped teams create software that aimed to support users at any point in the software.

### A. Threats to Validity

No empirical study is perfect. One reason is the inherent trade-off among different types of validity [51].

*External validity* refers to the ability to generalize findings. The fixes presented in this paper are context dependent on the teams and the software they evaluated, which means (1) the study might not be replicable and (2) the results may not be generalizable. We partially mitigate this risk by investigating inclusivity fixes across multiple studies and literature for triangulation in different software, but even so, these fixes may not be appropriate for some interfaces. For example, including step-by-step guidelines may be suitable when completing a procedural task like registering for a sports league. But, such guidelines might not work well in other types of software such as, software that supports open-ended problem solving where there is no "right" procedure, or even contradicting the purpose of a free-form software (e.g., adventure game).

*Internal validity* refers to how the study design can influence conclusions of the study. For example, Study 2 followed Action Research, so we did not attempt to control for teams' prior design practices or knowledge of gender issues; even had we wanted to, there is a lack of robust measurements for these. There were several factors that may have influenced what we observed, such as team members' prior experience with inspection methods and the make-up of the teams. Therefore, some of the interpretations we made from the data might be different had we studied different teams or software. This impacts what we observed in our results. To reduce effects of the threats above, we collected data from multiple teams and software projects, and made extensive use of triangulation across teams and with literature, as detailed in Table V and VI.

### VIII. CONCLUSION

This paper presents seven potential fixes to InfoProc and Learning style inclusivity bugs (three for InfoProc and four for Learning). The potential InfoProc fixes rested on letting the user decide when, how and how much information they want, rather than the system deciding how much they "should" have. The potential Learning fixes rested upon clarifying the workflow to support process-oriented learners, while also helping prevent errors for tinkerers.

As the seven potential fixes highlight, *inclusive* software is not about trading off one population for another—it is about supporting diverse cognitive styles in one interface so that diverse users and cognitive diversity itself can thrive. Not only will different users' cognitive styles vary from one another: a single user's cognitive style will vary from one *situation* to another, such as when someone is facing an imminent deadline vs. when they are not. Thus, making software more flexible to cognitive diversity helps not only multiple populations: it helps everyone.